# Green beam lines, a challenging concept

F. Osswald,[1,a)] E. Traykov,[1] T. Durand,[2,3] M. Heine,[1] J. Michaud,[4] JC. Thomas[5]

[1] IPHC, CNRS/IN2P3, Université de Strasbourg, 67200, Strasbourg, France
[2] SUBATECH, CNRS/IN2P3, IMT Atlantique, Université de Nantes, 44307, Nantes, France
[3] GIP ARRONAX, 44817, Saint-Herblain, France
[4] LP2IB, CNRS/IN2P3, 33170, Gradignan, France
[5] GANIL, CEA/DRF-CNRS/IN2P3, 14000, Caen, France

a)Author to whom correspondence should be addressed: francis.osswald@iphc.cnrs.fr.

## ABSTRACT

Due to increasing environmental and economic constraints, optimization of ion beam transport and equipment design becomes essential. The future should be equipped with planet-friendly facilities, that is, solutions that reduce environmental impact and improve economic competitiveness. The tendency to increase the intensity of the current and the power of the beams obliges us and brings us to new challenges. Installations tend to have larger dimensions with increased areas, volumes, weights and costs. A new ion beam transport prototype was developed and used as a test bed to identify key issues to reduce beam losses and preserve transverse phase-space distributions with large acceptance conditions.

## I. INTRODUCTION

A prototype of ion beam transport module has been developed at the Institut Pluridisciplinaire Hubert Curien (IPHC) with the aim of reducing beam losses and enabling beamlines with large acceptance according to current standards. This challenge is achievable by reducing emittance growth. The solution is planned for the transport of low energy ion beams but can be adapted to other conditions - it is based on electrostatic technology but can be converted into magnets. The prototype has a 0.5 m long quadripolar doublet structure and the design is based on high-order modes (HOM) analysis and pole shape and field maps optimization. The design will reduce optical aberrations produced by nonlinear fields that disturb the outer part of the beams (tail and halo), thereby improving the transport properties. The typical application is the transport of radioactive ion beams (RIBs), see SPIRAL 2 and DERICA projects for example.[1-3] Despite the low current intensity (< 10 nA) and standard loss level of $10^{-4}$, cumulative contamination can limit access and reduce operability.[4] There are many similar applications requiring enhanced beam transport, for example at SNS for the test facility,[5] at RIKEN for a focusing channel[6] used at LANL,[7] at IHEP for a multipole halo suppressor[8] and for Accelerator Driven Reactor Systems,[9] at FRIB for the ARIS magnetic separator,[10] at KIT[11] and FAIR[12] for storage rings with large acceptance, and at Daresbury for Laser Wakefield Acceleration.[13] Finally, the application for miniature accelerators is the new trend. The same design principle allowing large acceptance in the transverse trace-space (in position and angle) can also be used to reduce the equipment size, see Fig. 1 with two different bore radii and same beam.

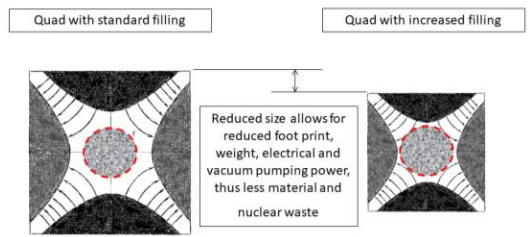

FIG. 1. Principle of the quadrupole with a large acceptance and reduction in size. Left: beam centered in a uniform field region, with standard filling (50-60%), reduced harmonic composition, and beam losses for long beam lines. Right: increased filling (80-90%), enlarged acceptance, non-uniform fields, presence of high-order harmonics, reduced emittance growth due to optimized pole shape, low beam losses, and reduced manufacturing and operation costs.

An experimental campaign conducted by IPHC and IN2P3 laboratories was carried out recently in order to characterize the performance of the module and identify the key problems related to the design of lossless beam transport and large acceptance lines. The results and the difficulties encountered in characterizing the performance of the doublet are reported with emphasis on low current intensity measurement (tail and halo of the beam), 2D transverse emittance figures with non-standard elliptical shape, background noise, control of experimental conditions such as beam stability, power supply regulation, and beam settings. The method to evaluate the beam transport performances with a filling of the quads ≥80% will be described as well as other side effects related to absolute emittance measurements.

## II. DESIGN OF A PROTOTYPE

### A. HOM analysis

The prototype is an optical structure based on an electrostatic quadrupole doublet. The design and manufacture of the prototype and especially the poles were carried out under a research contract between the CNRS and the Sigmaphi company in France. The design of the poles is based on an adequate ratio between the radius of the pole and the radius of the bore. It is carried out following an analysis of the fields and a decomposition by Fast Fourier Transform (FFT) of the higher-order multipole fields. The principle of the method consists in minimizing certain higher-order modes (HOM) at a certain radius and obtaining adequate beam optics after integrating the equations of motion of the particles along the length of the doublet. The iterative design of the poles is carried out with routines for parametric modeling and optimization of the OPERA3D code. Encouraging results have been obtained at TRIUMF for example in view of the design phase of ARIEL[14] (TRIUMF internal report, 1995), and precise models and a prototype have been produced at IPHC.[15]

After fabrication of the doublet, control of the resulting field in the aperture and near the surface of the quads is difficult due to the electrostatic influence of the probe and the desired precision. Since then, quality assurance based on a rigorous machining process and a 3D control procedure for parts and assemblies have been implemented. For example, a multiaxial arm was used to control the position of the assembled poles and the holding structure inside their vacuum chamber with a tolerance of ±100 µm (FARO equipment). Nevertheless, not everything designed can be achieved at reasonable costs with a conventional CNC machining center. Prospects for reducing costs (especially personnel costs due to production time) and allowing the manufacturing of complex pole shapes are emerging thanks to recent advances in the field of additive manufacturing, see for example.[16-17]

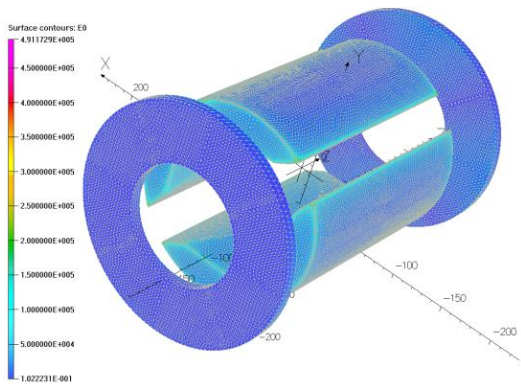

FIG.2. Numerical 3D model with field map on the surfaces of the quadrupole doublet and the two collimator rings.

### B. Beam optics

Mega science projects in the field of high energy and nuclear physics require long lines to transport the beam from the particle source to the target, so the control of beam losses is mandatory especially if the power of the stable beams or the RIBs intensity are high. The design of the prototype quadrupoles follows an iterative and interactive process with geometry variation, HOM field analysis for large radius i.e. large quad acceptance and filling, and beam optics calculation. A staged simulation approach is applied to three successive models: a single quad with ideal environment in order to evaluate the raw shape of the pole, a quad doublet positioned in the real environment with the mechanical structure and the vacuum chamber to obtain the fields and dynamics of the beam, and finally a virtual beam line with a fixed number of identical quad doublets in order to amplify the defects and observe detailed behavior of the optical structure over a long distance. Beam optics calculations are performed with TraceWin code.[18] The result of the transverse distributions in 2D phase-space for a 34 m long beam transport line is shown in Fig. 3. The optical structure is composed of six-unit cells of identical doublets. The initial Gaussian distributions have a marginal emittance of 80 π mm.mrad and are truncated at ±3 σ. With the optimized design the rms emittance growth is less than the usual 1% in both transverse planes (with a conventional design), the variation of the rms beam radii $\sigma_x$ and $\sigma_y$ is about 0.5% and the transmission is better than 99.5% thanks to a quad filling set at 80 %.[4] The filamentation of the emittances is due to the $3^{rd}$ order aberrations and the nonlinear fields applied to the particles approaching the poles. The effect is typical of transport lines but can be observed on periodic structures. The radial field increases towards the poles so that the outer particles no longer have the same path. The oscillations of the outer particles are non-harmonic while those near the center remain paraxial with a constant betatron wavelength.

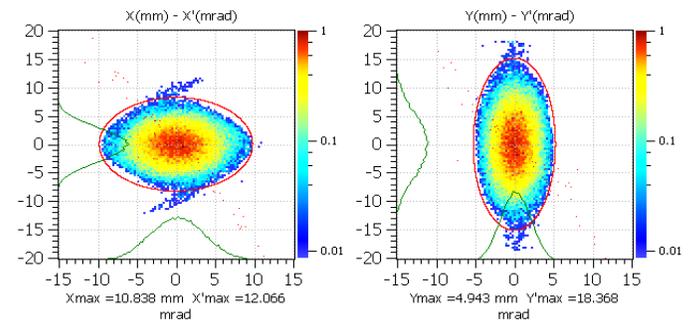

FIG. 3. Transverse trace-space distributions in X-X' and Y-Y' planes. The red ellipses indicate the rms emittances. The charged particles of the tail and the halo are submitted to nonlinear fields near the poles with spiral trajectories. These particles will be lost in the beam line after a certain distance.

## C. Design optimization

Pole design optimization is performed with objective functions and design variables in the Optimizer module of the OPERA3D code. The objective functions are defined by the quadrupole component (A2) which must be kept constant and the unwanted HOMs which must be minimized. Unwanted HOMs are defined by their negative impact on particle trajectories. For example, some components have a high-order radial dependence, with a radial field proportional to the radius to the fifth power and higher. It is a determining factor for particles passing close to the edges and for the beam halo which can be used as a signature of the performance (radial filling ≥80%). The technique is similar to the shimming performed for the cyclotron, separator, and spectrometer magnets. The entire development chain transferred from Sigmaphi was taken over at the IPHC institute. Precise termination criteria and numerical limits of design variables are defined for this purpose and avoid divergence during the iterations. Some specific routines have been developed since the first design based on hyperbolic poles, cut offs and fine shimming.[15] A new script has been defined recently to build a volume showing a smooth profile from "smoothing" operations based on sweeps, extrusions and cuts following a predetermined number and size of steps and within certain limits (the geometric constraints), see Fig. 4. The construction is controlled by the minimization of a few components, mainly the dodecapole (12 poles, component A6), the duodecapole (20 poles, component A10), and the 28 poles (component A14) because they tend to defocus the beam and lead to aberrations. Then, an experimentation plan is defined in order to optimize the search of the best solution in a minimum of time.

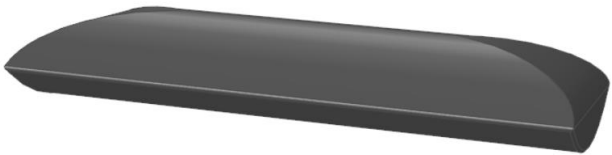

FIG. 4. Optimized pole design obtained from the original shape and additional geometric constraints leading to smoothing the profile and minimizing some harmonics. The shape is the result of an experimentation plan which takes 1 hour on a standard laptop (64 bits, 2.8 GHz, 32 Go RAM).

## III. EXPERIMENT

The objective of the experiment is to evaluate the emittance growth for different beam sizes, i.e. for several fillings of the first quadrupole aperture of the doublet (≥ 80 %). The differential measurements are carried out with two emittance scanners positioned respectively upstream and downstream of the prototype.

Measuring the transverse phase-space distributions is crucial and requires high resolution instruments in order to detect the tiny signature of the HOM field components on the beam tail and halo. The lower the disturbance of the initial emittance ellipse, the better the transport of the beam. Both emittance scanners were calibrated in a preliminary experiment.[19] The first device is a commercial 4D pepperpot system from Pantechnik, and the second is an "Allison" type 2D scanner developed at the IPHC.[20] The experiment is performed at the ARIBE-D4 line operated by the CIMAP laboratory and GANIL, Caen. The beamline delivers an analyzed $^{40}Ar^{8+}$ beam (without contaminants), and offers a variety of beam instrumentation equipment allowing for the different beam settings and the large beam transverse dimensions (up to 100 mm in diameter).

## A. Experimental setup

The test bench with the quadrupole doublet structure has been installed on the ARIBE line. The facility delivers a 15 keV/q $^{40}Ar^{8+}$ beam with an intensity typically between 100 nA and 1 µA DC. The beamline is equipped with an ECR ion source, several dipole (D), quadrupole (Q), and steering magnets (DC). The experimental setup and its location in the ARIBE line is shown in Fig. 5, including the main distances between the elements. The figure also displays an overlay of the combined TraceWin model of the ARIBE line and the experiment. The exact matching of the element positions in TraceWin with the real positions is essential for a correct comparison between the results of the simulations and the experimental data. Significant position and angular beam offsets were present after the extraction from the ECR ion source, which could not possible be fully corrected using existing steering elements in ARIBE.

In order to prepare the beam for the experiment the F22 – F24 slits were used to redefine and center the beam on axis in both planes before the D4 dipole. Beam centering was checked using the PR23, PR41, and PR42 beam profilers. The F22-F24 slits have been set to very small apertures in the horizontal plane in order to reduce the horizontal emittance, thus limiting the contribution to the measurements. This was necessary due to the use of the FV vertical slit and the EM1 and EM2 emittance meters which were installed to measure the vertical beam emittance only. The experimental setup including the vertical slit (FV), quadrupole doublet (Q1 and Q2), emittance meters (EM1 and EM2), and their relative positions are shown in Fig. 5.

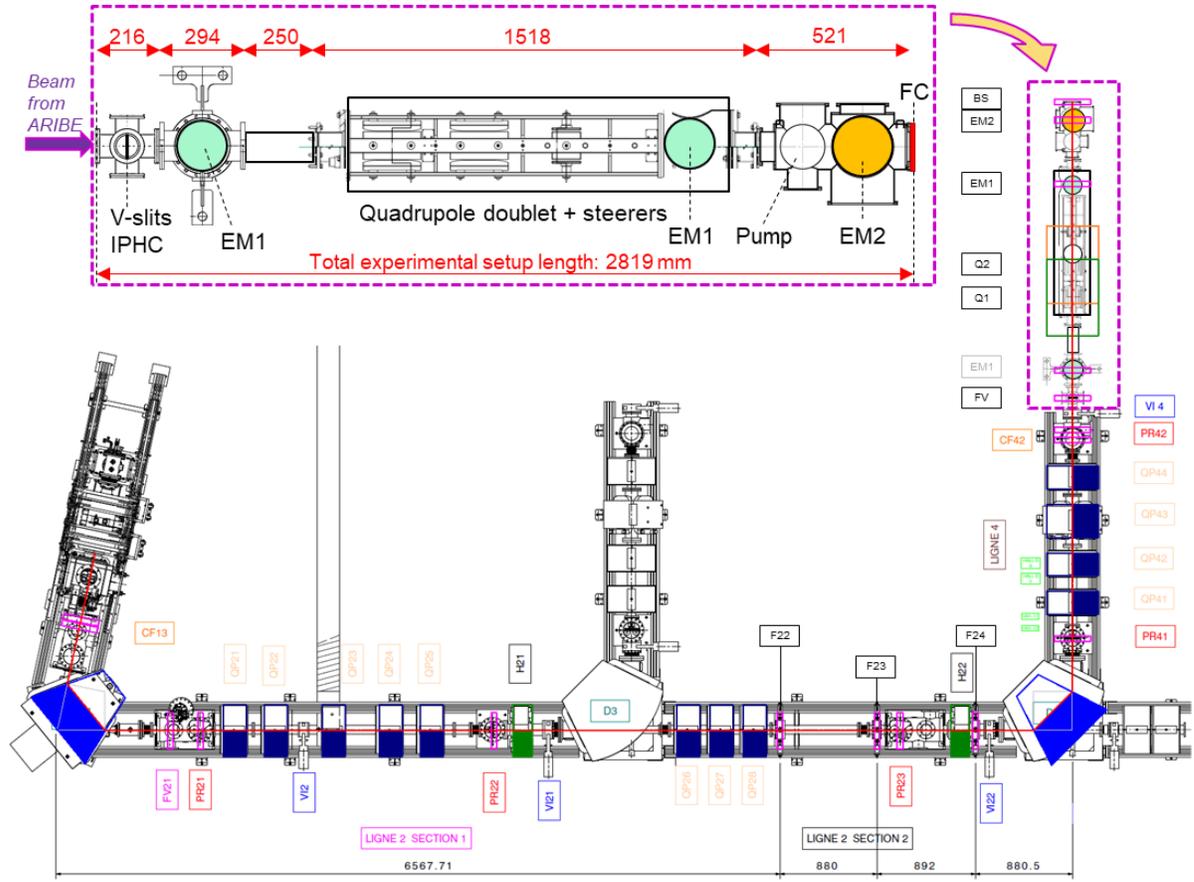

FIG.5. Experimental setup on the ARIBE-D4 beamline. The ECR ion source is positioned before the first dipole magnet (D1) on the left side of the beamline, followed by two dipole magnets (D3 and D4) and other equipment. The end of the line is identified by a vertical slit (FV) at the interface with the test bench (quad doublet optical structure). The FV slit enables to control the filling of the first quad of the doublet structure and therefore the large acceptance conditions of the experiment.

## B. Transverse distributions at different beam fillings

The beam filling is defined as the fraction of the aperture between the opposite poles at the entrance of the first quadrupole, i.e. ±50 mm at the Q1 input. It is estimated by two different methods. The first being purely geometrical assumes that a vertical beam waist is obtained at the position of PR42. This assumption is verified by the beam profile measured at PR42. The second method is based on the TraceWin simulations considering the exact tuning of the beamline. The simulation also confirms the vertical waist at PR42 as shown by the vertical envelopes in Fig. 6.

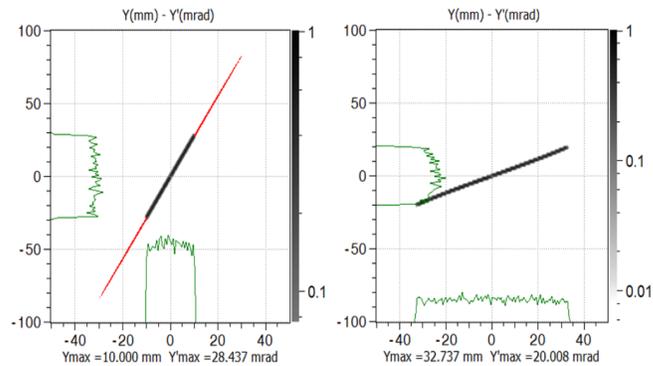

FIG. 7. Vertical distributions at the FV slit (at ±10 mm) and at the Q1 entrance (810 mm downstream) according to the simulations. Red points indicate the particles that have been intercepted by the slit.

The results for the beam fillings obtained from the two estimates are presented in Table 1 and in Fig. 8. The results obtained by the two different methods are very close except at very large slit apertures due to beam losses. The difference between the two series of measurements corresponds to a range between -2 and +15%. As shown in Fig. 8, the last point diverges due to the defined opening of ±50 mm in TraceWin, which is equivalent to 100% of the filling, whereas in the geometric estimation no limit was set by the opening, i.e. the beam could continue to grow beyond ±50 mm.

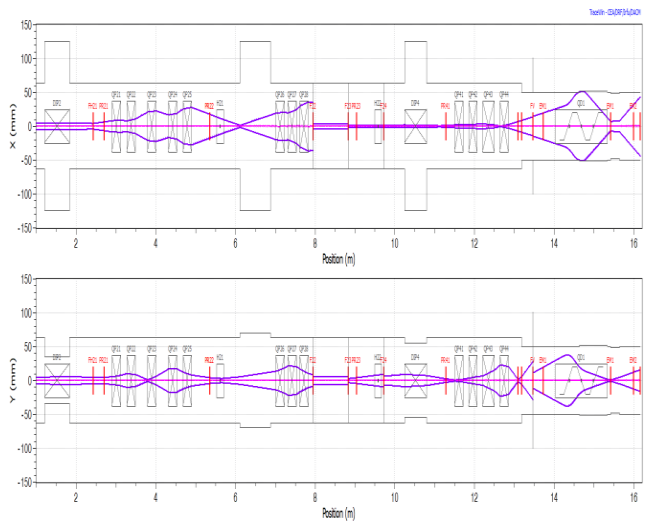

FIG. 6. Beam optics with rms envelopes in both transverse planes. Beam matching is mainly obtained with the quadrupoles (QP21-QP44) and the emittance matching section composed of three slits (F22-F24). The final adaptation is obtained with a vertical slit (FV) and a quadrupole doublet (QD1) in the analysis plane.

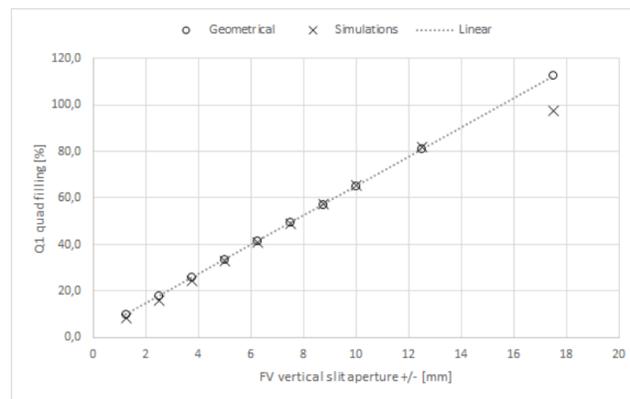

FIG. 8. Comparison of the filling at the first quad (Q1) obtained with the two different methods.

According to the simulations, the FV slit opening defines the beam filling at the Q1 entrance with good accuracy. Figure 7 shows the simulated vertical beam distributions at the slit and at the entrance of Q1. The slit is set to a ±10 mm opening corresponding to a vertical size of ±33 mm (FWHM) at Q1 and resembling a uniform distribution, thus allowing good definition of the filling.

TABLE I. Comparison of the filling of the first quad (Q1) obtained with two different methods.

| Slit opening ± [mm] | Geometrical Half-size at Q1 [mm] | Geometrical Filling [%] | Simulations Half-size at Q1 [mm] | Simulations Filling [%] |
|---|---|---|---|---|
| 1.25 | 4.94 | 9.9 | 4.2 | 8.4 |
| 2.5 | 8.88 | 17.8 | 8.1 | 16.2 |
| 3.75 | 12.82 | 25.6 | 12.2 | 24.4 |
| 5 | 16.75 | 33.5 | 16.4 | 32.8 |
| 6.25 | 20.69 | 41.4 | 20.5 | 41.0 |
| 7.5 | 24.63 | 49.3 | 24.5 | 49.0 |
| 8.75 | 28.57 | 57.1 | 28.6 | 57.2 |
| 10 | 32.51 | 65.0 | 32.8 | 65.6 |
| 12.5 | 40.38 | 80.8 | 41 | 82.0 |
| 17.5 | 56.14 | 112.3 | 48.7 | 97.4 |

## C. Emittance measurements

One of the main objectives of the experiment was to measure the performance of the quadrupole doublet (QD), that is, the emittance growth at different beam fillings. To measure the effects of filling, two different emittance meters (EM1 and EM2) were installed downstream of the QD. EM1 is the pepperpot system of Pantechnik. Measurements were set in the vertical plane to exclude unwanted high-order contributions from the dispersion of the upstream dipoles in the ARIBE line. Y-Y' distributions and extracted emittances measured with EM2 (and EM1) include the influence of higher order harmonics produced by the QD optical structure on the beam at different fillings. Simulation estimates show that, for a single QD unit, the higher order contribution and some emittance growth should be observable at ≥80% filling for a beam with uniform distribution. In order to have a good control of the filling and to form an almost uniform beam distribution, the beam was strongly diverged at the vertical slit (FV) installed at 810 mm upstream of the Q1 entrance. The beam divergence and the desired settings of the quadrupoles of the ARIBE-D4 line were defined using simulations. The obtained results from the EM2 measurements are shown in Fig. 9.

As expected, the beam emittance increases with slit opening. However, the unexpected observation was the large position and angular shift at EM2, and the optical aberrations on the emittance figures. This shift can be attributed to accidental internal electrical disconnections and/or charging of floating conductive surfaces inside the vacuum chamber. Indeed, during the measurements potential drops were observed on several of the HV power supplies (PS) of the QD prototype. These potential drops (a rather permanent drop of 10 to 40%) were associated with a large beam partially colliding with the poles and/or the internal structure of the setup. Another scenario that needs to be studied in more detail is the possibility of a more or less partial inter-pole discharge due to the large beam filling. The potentials could be tuned to their initial value after resetting the PS and containing the beam dimensions either by the use of slit and/or by an alternate tuning of the quadrupoles in ARIBE-D4. Then, asymmetry is visible from 30% filling and aberrations from 40%, see Fig. 9. These deleterious effects are related to the parasitic steering and the weak centering of the beam in the doublet structure. Another observation that could not be explained directly is the shape of the Y-Y' distribution at ±10 mm and ±12.5 mm. Additional beam structure can be observed at the lower right tail of the first, while such structure is not observed at the larger aperture. This may seem like a contradictory observation, but it can be explained by the sequence of the measurements, the timing of each of them is displayed in the plots of Fig. 9. It can be assumed that extensive charging of floating/disconnected internal part can lead to the formation of electric fields responsible for the parasitic deflection of the beam. This would explain the structure observed at ±10 mm as a contribution from the deflected beam, whereas the measurement at ±12.5 mm was made immediately after another measurement with the slit at ±1.25 mm. In the latter case, the beam emittance is reduced significantly, thus the ions cannot collide and charge the surrounding surface even in the presence of large angular and spatial offsets.

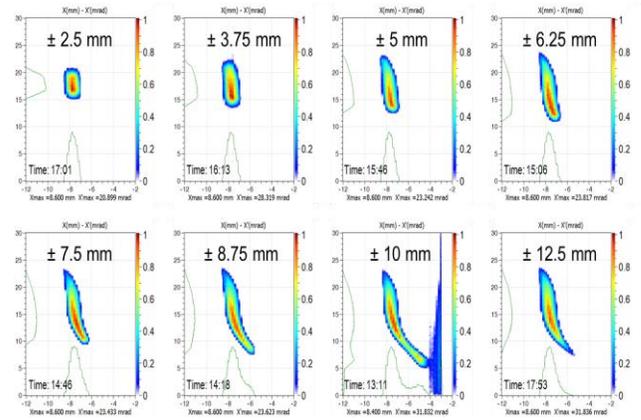

FIG. 9. Emittance measurements in 2D with different slit apertures (± half opening of FV slit).

## D. Measurement of the beam profiles and losses

The progressive opening of the slit (FV) and the beam currents read on their jaws and on the beam stop (BS) make it possible to reconstitute the beam profile at the location of the slit. The profile was determined independently by the two currents and the results are shown in Fig. 10. During beam filling measurements, the beam current was read from both the jaws of the slit (FV) and the beam stop (BS) installed on the blind flange at the end of the setup. Normalization of the measurements was necessary due to the difference in material/finish of the slit jaws and FC and due to the losses in the horizontal plane associated with the smaller diameter of the beam stop compared to the horizontal dimension of the beam. A scale factor of 1.82 was applied to the BS reading to keep the sum of the two currents constant for small slit openings.

The matching between the two curves (based on the two readout currents) is very good at low apertures and only diverges at very large apertures due to the losses associated with the large filling. Indeed, at small apertures, the transmission of the beam passing through the QD structure is maximum, whereas at larger apertures (i.e. larger fillings), a fraction of the beam collides with the poles due to the spurious steering, resulting in the observed losses, as shown in Fig. 11. One should note that the two data points that deviate from the curve are the same in Fig. 10 and 11.

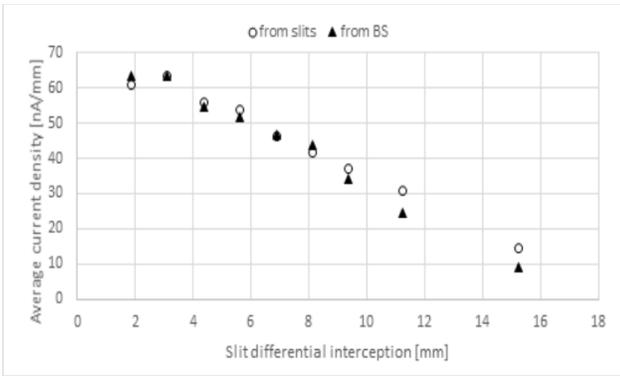

FIG. 10. Profile reconstruction from vertical slit (FV) and beam stop (BS) measurements. The discrepancy between the two series is due to beam losses occurring after the slit.

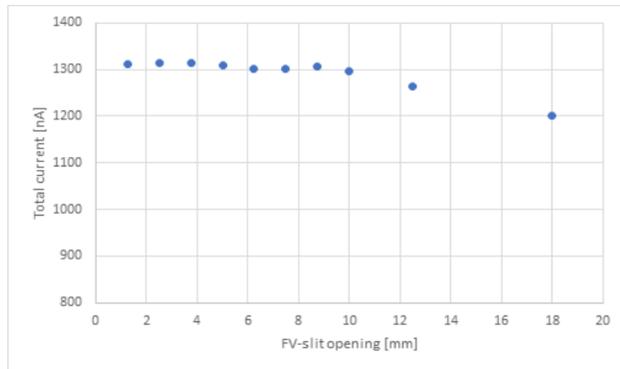

FIG.11. Total beam current intensity as a function of vertical slit opening (FV). The slope of the curve is produced by the beam losses as mentioned in Fig. 10.

## IV. DISCUSSION

The main difficulty in evaluating the performance of the quadrupole doublet was related to the measurement of the signature of the HOM components in the tail and the halo of the beam and to detect the emittance growth. Accuracy is limited by the spatial and angular resolution of emittance scanners and, as current intensities were limited to 1pA during the experiment for both units, this barrier should clearly be pushed back in the future, see Fig. 12.

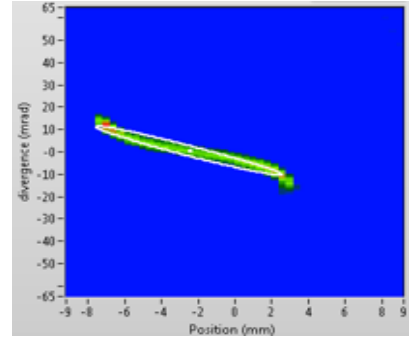

FIG. 12. Distribution of trace-space in the Y-Y' vertical plane obtained with the pepperpot scanner (EM1), slit opening of ±12.5 mm (FV), and 80% filling in the first quad (Q1 of QD quadrupolar doublet). Most of the beam tail and of the halo disappears due to high threshold settings and background noise filtering.

The halo is characterized by a low charge density around the core and the tail of the beam. Some typical numbers for the halo are <10% of the total particle count, >60% of the area of the 2D beam emittance, <$10^{-5}$ of the total beam current intensity at 5 σ for a Gaussian distribution, see references.[5,21,22] During the experiment, the current intensity of the single charged argon beam was typically 100 nA. The beamlet resulting of the scanner sampling has a current intensity of 1 nA and the standard resolution of the electronics is 10 pA. This leads to a minimum/total intensity ratio of $10^{-4}$ reflecting the problem of measuring the emittance at 4 or 5σ ($10^{-5}$ is required). When measuring such low intensities, the signal to noise ratio (S/N) becomes crucial and therefore the identification of the background noise (BGN), the filtering of the signal, and the reconstruction of the emittance figure keeping most of the halo will require further effort to improve emittance growth measurements. A specific algorithm was developed for the experiment to deal with non-Gaussian beam distributions and noisy measurements. Based on 2D-Gaussian and polynomial fits, it has significantly improved the accuracy of emittance measurements performed with the IPHC scanner.[19] Alternative developments include the definition of the background based on the statistical fluctuations far from the expected signal, which allows to decompose the entire emittance figure using non-analytical functions (spline fits) without constraints on the area where potential tails lie. It is foreseen to make use of artificial intelligence algorithms for image recognition of characteristic emittance shapes in the initial step and during the definition of tails and halos.

As noted earlier in a reference,[23] there are some inherent difficulties in measuring emittance in low-energy beam transport lines which are related to contaminants, space charge, optical aberrations, non-Gaussian distributions, and non-elliptical emittance figures which induce errors in the representation of the trace-space and for the rms calculation. Accurate measurement of absolute emittance is a difficult task and comparison between two different systems is even more difficult. A 10-20% discrepancy between two different systems is common. Therefore, the emittance growth must be measured with the same device with successive focused and defocused beam, even if a comparison with a second system placed under the same conditions, placed nearby, and during the same run is strongly recommended, see EM2 emittances in Fig. 13.

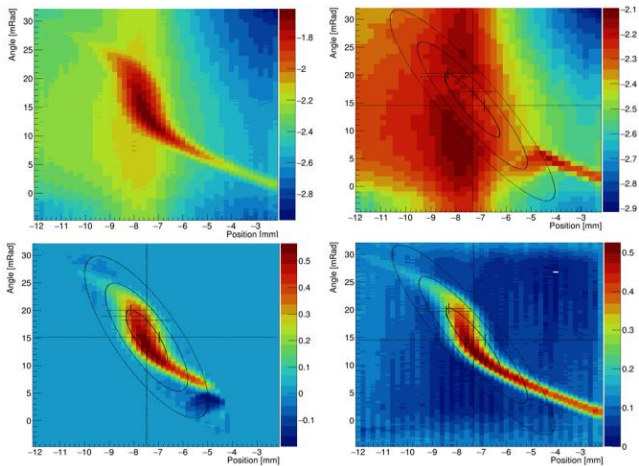

FIG. 13. From top left to bottom right: (a) raw data, (b) background noise identification (BGN), (c) BGN filtering, and (d) emittance figure reconstruction highlighting tails produced by optical aberrations. The measurements are performed with the same conditions than in Fig. 12 i.e. with 80% filling in Q1.

Other solutions are to be developed in order to allow the characterization of the prototype performance: the stability of the PS must be improved. Although current PS specifications indicate a high stability and $10^{-5}$ load regulation, behavior under load variation, with current injection into the electric circuit, induced voltage transients, and the use of a resistive load (bleeder resistor) must be investigated in more detail. This will allow control of the pole potential even with beam losses and inter-pole effects.

## V. CONCLUSION

A low-energy ion beam transport prototype including a quadrupole doublet has been realized after optimization of the quadrupole design and analysis of electrical field harmonics. From simulations of the beam dynamics carried out on a unitary optical cell composed of a quadrupole doublet as well as on a virtual line composed of six identical doublets, it was observed that the minimization of some of the HOM components results in an attenuation of the optical aberrations and therefore to a reduction of the beam losses. The design can be further improved by new optimization routines developed since the first investigations. Moreover, additive manufacturing could open up new perspectives to reduce costs and enable complex pole shapes for quadrupoles. The experiment carried out with the doublet structure on an ion beam line has revealed the limits of the measurements of the trace-space currently possible and necessary to evaluate the tiny emittance growth generated by a unit optical cell. The importance of the signal analysis and image reconstruction correlated with background noise, non-Gaussian distributions, and non-elliptical emittance figures was demonstrated because classical statistical tools were not entirely satisfactory. Problems related to beam settings were highlighted (stability, control of the dimensions, reproducibility, uniform transverse distribution, quadrupole filling, etc.) as well as the importance of the reliability of the electric power supply (influence of the beam on the poles, beam losses, electrical discharges, and induced transients).

## VI. ACKNOWLEDGMENTS

We acknowledge the experiment P1246-A granted by the GANIL iPAC and the ARIBE facility that are jointly run by the GANIL and CIMAP laboratories. We also acknowledge the 300-kV implanter of the ACACIA facility at iCUBE laboratory, CNRS/INP Strasbourg, which was used for commissioning and testing some equipment.

F.O. wishes to thank I. Ivanenko, JINR, and W. Beeckman, Sigmaphi, for the delivery of accurate 3D field maps and fruitful discussions.

The data analysis was partly supported by grants from the French National Agency for Research, ARRONAX-Plus n°ANR-11-EQPX-0004, IRON n°ANR-11-LABX-18-01, and ISITE NExt no ANR-16-IDE-0007. It is supported by a PhD scholarship from the Institute of Nuclear and Particle Physics (IN2P3) from the National Scientific Research Center (CNRS).

## AUTHOR CONTRIBUTIONS

Teddy Durand: formal analysis, software, visualization, and validation. Marcel Heine: formal analysis, writing/review and editing. Julien Michaud: formal analysis, software, visualization, and validation. Francis Osswald: supervision conceptualization, methodology, validation, writing/original draft preparation, writing/review, and editing. Jean-Charles Thomas: formal analysis and validation. Emil Traykov: conceptualization, formal analysis, methodology, software, validation, visualization, and writing/original draft preparation.

## AUTHOR DECLARATIONS

### Conflict of Interest
The author has no conflicts to disclose.

## DATA AVAILABILITY

The data that support the findings of this study are available from the corresponding author upon reasonable request.